%% file: manuscript.tex
\documentclass[a4paper,fleqn,useAMS,usenatbib]{mnras}

\pdfoutput=1

\usepackage{mathptmx}

\usepackage[T1]{fontenc}
\usepackage{ae,aecompl}

\usepackage[dvips]{graphicx}
\usepackage{amsmath}
\usepackage{amsfonts}
\usepackage{amssymb}
\usepackage{comment}
\usepackage[dvipsnames]{xcolor}
\usepackage[%
        ]{hyperref}
        
\hypersetup{
	colorlinks=true,	
	urlcolor=MidnightBlue,		
	pdfpagelabels=true,
	hypertexnames=true,
	plainpages=false,
	naturalnames=true,
	pdftitle={A resonant pair of warm giant planets revealed by TESS},    
	pdfauthor={Kipping et al.},     
	linkcolor=WildStrawberry,          
	citecolor=ForestGreen,        
}

\input{shortcuts.tex}

\title[A resonant pair of warm giant planets revealed by \TESS]{
A resonant pair of warm giant planets revealed by \TESS
}
\author[Kipping et al.]{David Kipping$^{1}$\thanks{E-mail:
dkipping@astro.columbia.edu},
David Nesvorn\'y$^2$,
Joel Hartman$^3$,
Guillermo Torres$^4$,
\newauthor{Gaspar Bakos$^3$, Tiffany Jansen$^1$, Alex Teachey$^1$}
\\
$^1$Dept. of Astronomy, Columbia University, 550 W 120th Street, New York NY 10027, USA\\
$^2$Dept. of Space Studies, Southwest Research Institute, 1050 Walnut Street, Suite 300, Boulder, CO 80302, USA\\
$^3$Dept. of Astrophysical Sciences, Princeton University, NJ 08544, USA\\
$^4$Harvard-Smithsonian Center for Astrophysics, 60 Garden Street, Cambridge, MA 02138, USA \\
}

\date{Accepted . Received ; in original form }

\pubyear{2019}

\begin{document}
\label{firstpage}
\pagerange{\pageref{firstpage}--\pageref{lastpage}}
\maketitle

\begin{abstract}
We present the discovery of a pair of transiting giant planets using
four sectors of \TESS\ photometry. TOI-216 is a $0.87$\,$M_{\odot}$ dwarf orbited
by two transiters with radii of $8.2$\,$R_{\oplus}$ and $11.3$\,$R_{\oplus}$,
and periods of $17.01$\,d and $34.57$\,d, respectively. Anti-correlated TTVs are
clearly evident indicating that the transiters orbit the same star and interact
via a near 2:1 mean motion resonance. By fitting the TTVs with a dynamical model, we infer
masses of $30_{-14}^{+20}$\,$M_{\oplus}$ and $200_{-100}^{+170}$\,$M_{\oplus}$,
establishing that the objects are planetary in nature and have likely sub-Kronian
and Kronian densities. TOI-216 lies close to the southern ecliptic pole and
thus will be observed by \TESS\ throughout the first year, providing an
opportunity for continuous dynamical monitoring and considerable
refinement of the dynamical masses presented here. TOI-216 closely resembles
Kepler-9 in architecture, and we hypothesize that in such systems these
Saturn-analogs failed to fully open a gap and thus migrated far deeper into the
system before becoming trapped into resonance, which would imply that future detections of
new analogs may also have sub-Jupiter masses.
\end{abstract}


\section{Introduction}
\label{sec:intro}

\input{introduction.tex}

\section{Observations}
\label{sec:observations}

\input{observations.tex}

\section{Analysis}
\label{sec:analysis}

\input{analysis.tex}

\section{Discussion}
\label{sec:discussion}

\input{discussion.tex}
	
\section*{Acknowledgments}

DMK is supported by the Alfred P. Sloan Foundation Fellowship.
DN's work was supported by the NASA Exoplanet Research Program (XRP).
AT \& TJ are supported through the NSF Graduate Research Fellowship (DGE 16-44869).

Funding for the TESS mission is provided by NASA's Science Mission directorate. We acknowledge the use of public TESS Alert data from pipelines at the TESS Science Office and at the TESS Science Processing Operations Center. This paper includes data collected by the TESS mission, which are publicly available from the Mikulski Archive for Space Telescopes (MAST).

This work has made use of data from the European Space Agency (ESA) mission Gaia (https://www.cosmos.esa.int/ gaia), processed by the Gaia Data Processing and Analysis Consortium (DPAC, https://www.cosmos.esa.int/web/gaia/ dpac/consortium). Funding for the DPAC has been provided by national institutions, in particular the institutions partic- ipating in the Gaia Multilateral Agreement. This research has made use of NASA's Astrophysics Data System. This research has made use of the SIMBAD database, operated at CDS, Strasbourg, France. This research has made use of the NASA Exoplanet Archive and the Exoplanet Follow-up Observation Program website, which are operated by the California Institute of Technology, under contract with the National Aeronautics and Space Administration under the Exoplanet Exploration Program.

\textit{Facilities:} TESS

\textit{Software:} \multi\ \citep{feroz:2008,feroz:2009},
\isochrones\ \citep{morton:2015}, \mercury\ \citep{chambers:1999},
{\tt corner.py} \citep{corner}


%

\bsp
\label{lastpage}
\end{document}

%% file: shortcuts.tex
\newcommand{\TESS}{{\it TESS}}
\newcommand{\Gaia}{{\it Gaia}}
\newcommand{\Kepler}{{\it Kepler}}

\newcommand{\isochrones}{{\tt isochrones}}
\newcommand{\dartmouth}{{\tt Dartmouth}}
\newcommand{\mercury}{{\tt Mercury6}}
\newcommand{\TIC}{TIC 55652896}
\newcommand{\cofiam}{{\tt CoFiAM}}
\newcommand{\polyam}{{\tt PolyAM}}
\newcommand{\innerp}{\mathrm{inner}}
\newcommand{\outerp}{\mathrm{outer}}
\newcommand{\multi}{{\sc MultiNest}}
\newcommand{\ourlink}{\href{https://github.com/CoolWorlds/TOI216}{this URL}}

%% file: introduction.tex
A small fraction of exoplanets in the cosmos have the correct orbital geometry
to transit their star as seen from our home. These transiting planets have
been a ``royal road to success'' in planet discovery
\citep{russell:1948,winn:2010} yielding thousands of discoveries in recent
years (see the NASA Exoplanet Archive; \citealt{akeson:2013}), despite the
fact that they represent just a sliver of the total population. An even more
rarefied population is that of transiting planets exhibiting transit timing
variations (TTVs; \citealt{agol:2005,holman:2005,deck:2015}).
For these special worlds, not only can one infer the planetary size from the
transit depths, but dynamical modeling of the TTVs can often provide
planetary masses too - a fact heavily exploited by \Kepler\ \citep{holman:2010,
lithwick:2012,nesvorny:2012}. In such cases, it is therefore possible to
confirm the planetary nature of a system almost exclusively from photometric
observations (e.g. \citealt{ford:2011,steffen:2013}).

\Kepler\ enjoyed many successes with this strategy, largely enabled by its
patience to stare at the same stars for over four years continuously. With
\TESS, the full-sky nature of the survey means that most parts of the sky are
observed for much shorter windows\footnote{This situation could change if
with an extended \TESS\ mission \citep{bouma:2017}}, potentially posing a challenge to dynamical
confirmation of planetary candidates. However, \TESS\ does maintain a longer
vigil on the ecliptic poles, observing these fields for up to a year
continuously \citep{ricker:2016}.

In this work, we describe the discovery of two \TESS\ planets near a
2:1 mean motion resonance (MMR) leading to highly significant TTVs. Thanks to
the host star's fortitious location near the southern ecliptic pole, \TESS\
can observe the target for most of the first year making TOI-216 an
excellent target for monitoring planet-planet interactions. We describe
the observations by \TESS\ in Section~\ref{sec:observations} with
attention to detrending, contamination and stellar properties. In
Section~\ref{sec:analysis} we regress light curve models and TTV models,
demonstrating that the system is a pair of planet-mass objects
gravitationally interacting with one another. Finally, we discuss the
possibilities opened up by this exciting new system in
Section~\ref{sec:discussion}.

%% file: observations.tex
\subsection{Identification}
\label{sub:identification}

\TIC\ was observed by \TESS\ in the first four sectors of year one, and indeed
is scheduled to be observed in every sector of that year except for sector 10
(2019-Mar-26 to 2019-Apr-22). Falling on camera 4, the target is a relatively
rare example of a transiting planetary system caught within the \TESS\ continuous
viewing zone (CVZ). With an ecliptic latitude of -82.476408$^{\circ}$, we
highlight that the target also lands close to JWST's planned CVZ and would
be observable for $\gtrsim 260$\,days per year of the mission.

A \TESS\ alert was issued on 2018-11-30 for two candidate transiting planets
associated with \TIC\ using sectors 1 and 2, dubbed TOI-216.01 and TOI-216.02.
With periods of $\sim 17.1$\,days (TOI-216.02) and $\sim 34.5$\,days
(TOI-216.01), the outer candidate was only seen to transit twice during this
time (once per sector).

Amongst the 300+ TOIs identified at the time of writing, this pair stood out
as particularly interesting because the planetary candidates lie near a 2:1
period commensurability. If the objects were orbiting the same star, and
gravitationally interacting, then it may be possible to confirm the planetary
nature of the pair without any ground based follow-up (e.g. see
\citealt{steffen:2013}). For this reason, we decided to further study this
system.

\subsection{Stellar properties}
\label{sub:star}

TOI-216 has an apparent magnitude of 11.5 in the \TESS\ bandpass. From the
\TESS\ Input Catalog (TIC) version 7 \citep{stassun:2018}, catalog survey
spectroscopy of the star constrains $T_{\mathrm{eff}} = (5026\pm125)$\,K,
[M/H]$=0.32\pm0.10$ and $\log(g) = 4.66\pm0.20$. These properties are used by
TIC to infer $M_{\star} = (0.879\pm0.073)$\,$M_{\odot}$ and
$R_{\star}=(0.715\pm0.166)$\,$R_{\odot}$.

We also queried the star within \Gaia\ DR2 \citep{GAIADR2:2018} and
find a parallax measurement of $5.591\pm0.028$\,mas (GAIA DR2
4664811297844004352). Rather than use the TIC-7 summary statistics for stellar
mass and radius, we would prefer to work with posterior samples - as well as
include the \Gaia\ parallax - and so we elected to perform our own Bayesian
isochrone fitting. To this end, we used the \isochrones\ package by T. Morton
using the previously listed constraints on $V$, $T_{\mathrm{eff}}$, [M/H],
$\log g$ and parallax with the \dartmouth\ stellar evolution models.

\input{star_table.tex}

The inputs to our fits (the ``star.ini'' file) are given in the top
panel of Table~\ref{tab:star}, and the derived parameters of interest
in the lower panel. As expected, our results closely agree with those listed
in the TIC, although our inference is more precise as a result of
using the GAIA parallax which was not available when TIC-7 was compiled.

\subsection{Contamination}
\label{sub:contamination}

With a pixel size of 21 arcseconds, there is a greater chance of crowding with
\TESS\ than \Kepler. The aperture used in each sector varies slightly but
is approximately 4 by 3 pixels and thus sources out to 84 arcseconds can
contaminate the aperture.

Fortunately, there are no comparably bright stars that lie within this region.
The nearest star listed in the TIC to our target is TIC 55652894, separated by
48.2 arcseconds but far fainter with an apparent \TESS\ magnitude of 16.3
(1.2\% the brightness level). \Gaia\ DR2 \citep{GAIADR2:2018} reports 46
stars within 84\,arcseconds, with $G$-band magnitudes from 17.1 to 21.1
(TOI-216 is 12.2). Together, these sources could maximally dilute the
target by 5.6\% in $G$, although the true value will be less due to color
correction to the redder \TESS\ bandpass, location of the sources and the
finite PSF widths.

These contaminating source cumulatively lead to a small amount of dilution of 
TOI-216, which is estimated within the \TESS\ aperture to be 0.33\%, 0.46\%,
0.29\%, 0.26\%, 0.58\% and 0.31\% for sectors 1 to 6 respectively (values taken
form the \TESS\ light curve files) and these are included in our later light
curve fits using the prescription of \citet{tinetti:2010}.

We also highlight that an unresolved companion may reveal itself through
centroid shifts during the moments of transit, but all centroid shifts for this
star are below 1\,$\sigma$ as reported by the \TESS\ sector 1-3 cumulative Data
Validation (DV) report.

\subsection{Light curve detrending}
\label{sub:detrending}

\begin{figure*}
\begin{center}
\includegraphics[width=2.05\columnwidth,angle=0,clip=true]{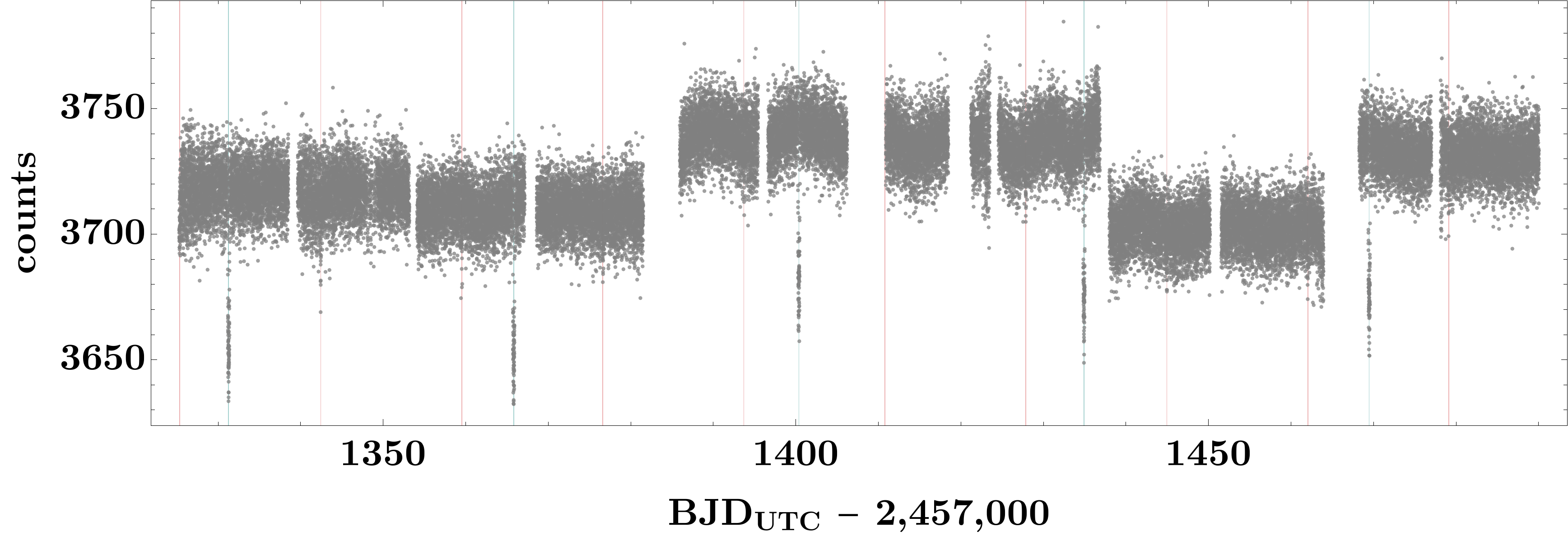}
\caption{\emph{
PDC light curve of \TIC\ as observed by \TESS\ for sectors 1 to 6. We mark
the location of the transits of TOI-216.01 in blue and TOI-216.02 in red.
}}
\label{fig:photometry}
\end{center}
\end{figure*}

We downloaded sectors 1 to 6 short-cadence (2\,minute) data for \TIC\ and work
with the PDC (Pre-search Data Conditioning) product in what follows
\citep{jenkins:2017}. Any bad data flags were removed, and outliers filtered
with 5\,$\sigma$ clipping against a 20-point moving median. Using the ephemeris
and duration for TOI-216.01 and TOI-216.02, we de-weight the in-transit points
for the purposes of detrending. The PDC light curve for all six sectors is
shown in Figure~\ref{fig:photometry}.

We next detrend each transit epoch of each planet independently, using four
different algorithms following \citet{TK18} - \cofiam, \polyam, local
polynomials and a Gaussian Process using a squared exponential kernel. The four
light curves are then combined into a single time series - a method
marginalized light curve - by taking the median at each time stamp and
inflating the formal uncertainties by adding the inter-method standard
deviation in quadrature. We direct the reader to \citet{TK18} for a detailed
description of the four algorithms as well as the method marginalization
process. The resulting light curves from all four methods, as well as the
method marginalized light curves, are made available at \ourlink.

We find that the inter-method standard deviation is many times smaller than
the formal uncertainties, indicating a highly stable detrending. The
median formal uncertainty is 2430\,ppm but the median inter-method standard
deviation is $16$ times smaller at 150\,ppm. After adding this extra component
in quadrature to the formal uncertainties, the error increases by just 0.2\%.

%% file: star_table.tex
\begin{table}
\caption{
\emph{
Medians and one-sigma uncertainties for the stellar properties of \TIC.
The top panel are the properties listed in the TIC version 6
\citep{stassun:2018}, as well as the \Gaia\ DR2 parallax. The
lower panel lists the derived properties through isochrone matching.
}
}
\centering 
\begin{tabular}{ll} 
\hline
parameter & value \\ [0.5ex] 
\hline
$V$ & $12.324 \pm 0.069$ \\ 
$T_{\mathrm{eff}}$\,K & $5026 \pm 125$ \\
$[M/H]$\,[dex] & $0.32 \pm 0.10$ \\
$\log g$\,[dex] & $4.66 \pm 0.20$ \\
parallax\,[mas] & $5.591 \pm 0.028$ \\
\hline
$M_{\star}$\,[$M_{\odot}$] & $0.874_{-0.034}^{+0.035}$ \\
$R_{\star}$\,[$R_{\odot}$] & $0.838_{-0.030}^{+0.043}$ \\
$\rho_{\star}$\,[kg\,m$^{-3}$] & $2090_{-300}^{+270}$ \\ 
$d$\,[pc] & $178.84_{-0.88}^{+0.89}$ \\ [1ex]
\hline 
\label{tab:star}
\end{tabular}
\end{table}

%% file: analysis.tex
\subsection{Light curve model}
\label{sub:model}

We initially built light curve models for the system which treated each
planetary candidate as orbiting an independent star. It became immediately
clear that the transiters displayed strong transit timing variations (TTVs),
as can be seen by simple inspection of Figure~\ref{fig:lightcurves}. More
detailed analysis of these light curves presented in Section~\ref{sub:TTVs}
reveals strong evidence for anti-correlation - the hallmark of dynamically
interacting planets \citep{steffen:2013}. This is only possible if the
two transiters are orbiting the same primary, and thus in our final light curve
modeling we decided to treat the objects as sharing a common host star.

This is particularly useful for modeling the inner candidate, TOI-216.02, whose
light curve displays a V-shaped morphology consistent with a grazing geometry.
Treated as an independent body, V-shaped transits display strong degeneracies
between size, impact parameter, limb darkening coefficients and host star density
\citep{carter:2008}. Since the anti-correlated TTVs imply a common host star,
and the other transiter is non-grazing, the conditional relationship greatly
aids in the inference of a unique light curve solution for TOI-216.02.

Our light curve model is that of the classic \citet{mandel:2002} quadratic limb
darkening code, which is oversampled by a factor of 5 to correct for the
slight distortion of finite integration time via the prescription of
\citet{binning:2010}. The quadratic limb darkening coefficients are re-parameterized
to $q_1$ \& $q_2$ following \citet{kipping:2010}.

Since TTVs are apparent in the light curve (Figure~\ref{fig:lightcurves}), we
allow each transit epoch to have a unique time of transit minimum, $\tau$. In
what follows, we also assume both transiters have nearly circular orbits
($e\simeq0$). If one (or both) were in fact eccentric, the derived stellar
density would be erroneous (see \citealt{kipping:2010,moorhead:2011,
tingley:2011,dawson:2012}) by a factor of (marginalizing over argument of
periastron, $\omega$):

\begin{align}
\frac{<\rho_{\star}'>}{\rho_{\star}} &= \frac{1}{2\pi} \int_{\omega=0}^{2\pi} \frac{(1+e\sin\omega)^3}{(1-e^2)^{3/2}} \mathrm{d}\omega,\nonumber\\
\qquad&= \frac{ 1+\tfrac{3}{2} e^2 }{ (1-e^2)^{3/2} }.
\label{eqn:photoeccentric}
\end{align}

Our later fits (see Section~\ref{sub:fits}) reveal that our light curve derived
density is measured to a precision of 6\% and thus from
Equation~(\ref{eqn:photoeccentric}) one may show that $e<0.14$ should be
expected to lead to a less than 1\,$\sigma$ systematic error in the inferred
density. As an apparently fairly compact, multiple planet system, we consider
this assumption is reasonable on the grounds of dynamical stability, and indeed
our later TTV fits favor low eccentricities (see Section~\ref{sub:TTVs}).
Nevertheless, we choose not to use the light curve derived density in any
attempt to refine the isochrone modeling from Section~\ref{sub:star}.

\subsection{Light curve fits}
\label{sub:fits}

Our light curve model has a total of 17 free parameters: two ratio-of-radii
($p_{\innerp}$ \& $p_{\outerp}$), two impact parameters ($b_{\innerp}$ \&
$b_{\outerp}$), two limb darkening coefficients ($q_1$ \& $q_2$), a mean
stellar density ($\rho_{\star}$), six times of transit minimum for the inner
transiter ($\tau_{\innerp,i}$) and four times of transit minimum for the outer
transiter ($\tau_{\outerp,i}$).
We assume uniform priors for all parameters except for $\rho_{\star}$ for which
we adopt a broad log-uniform prior. All of the priors are listed in
Table~\ref{tab:priors}

\input{priors_table.tex}

\begin{figure*}
\begin{center}
\includegraphics[width=2.05\columnwidth,angle=0,clip=true]{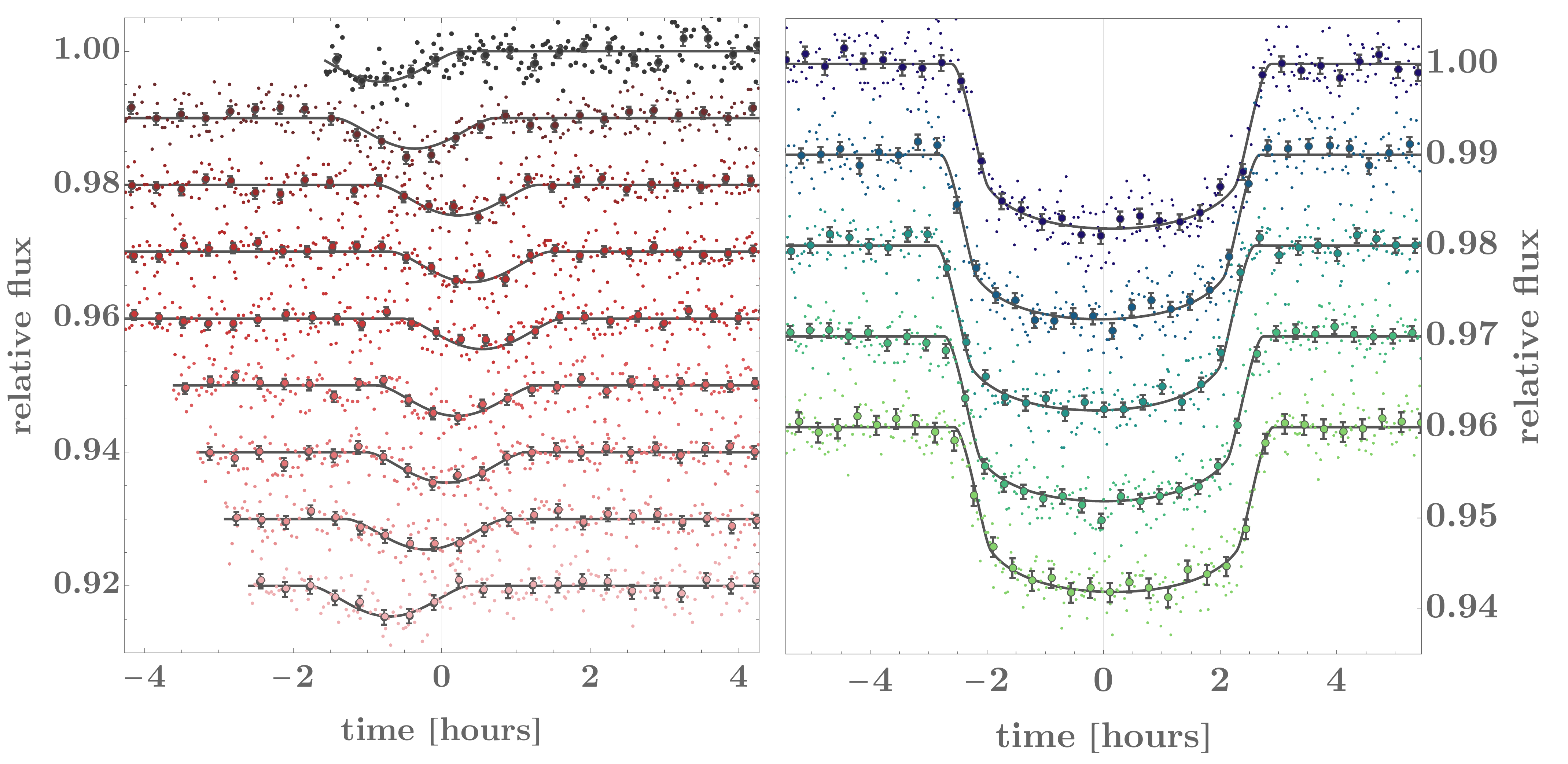}
\caption{\emph{
Left: The six available transits of TOI-216.02 observed by \TESS\ in sectors
1 to 4 phase folded on the best-fitting linear ephemeris. We bin the data to
20\,minute samples and overlay the maximum a-posteriori light curve model.
Right: Same, but for TOI-216.01, where only four transit are available. In
both cases, TTVs are clearly evident.
}}
\label{fig:lightcurves}
\end{center}
\end{figure*}

Fits were conducted using \multi\ \citep{feroz:2008,feroz:2009} with 4000 live
points\footnote{The recommended value is 2000 for posterior estimation (F.
Feroz; private communication), but we decided to double this to decrease the
chance of a missed mode.}, an evidence tolerance of 1.0 and in non-constant
efficiency mode. The maximum a-posteriori light curve
solution is plotted in Figure~\ref{fig:lightcurves}. We make the full posterior
samples available at \ourlink\ but list the credible intervals on the 10
transit times in Table~\ref{tab:times} and the other 7 global parameters in
Table~\ref{tab:final}.

\input{times_table.tex}

Considering the 7 global parameters, there are two noteworthy conclusions that
can be drawn from the results. The first is that the impact parameter of
the inner planet is unusually high at $b_{\innerp}=0.957_{-0.022}^{+0.047}$.
Given that the ratio-of-radii is measured to be
$p_{\innerp}=0.089_{-0.012}^{+0.032}$, then we have $b_{\innerp}>1-p_{\innerp}$
and thus this is a definitively grazing transit. Such transits are rare and have been
hypothesized to be powerful probes of nodal variations \citep{kipping:2009b},
and thus TOI-216.02 should be carefully monitored in the future for such
changes.

Second, the a-posteriori mean stellar density is found to be
$2380_{-140}^{+100}$\,kg\,m$^{-3}$. We remind the reader that
this was using a log-uniform prior and thus was inferred agnostically. The only
assumption in the model is that both transiters orbit the same star, which is
established from the anti-correlated TTVs, and that the eccentricities are
small ($e \lesssim 0.14$), which is reasonable given the system's compactness
and multiplicity for orbital stability. This density is consistent with
the independently derived value from our earlier isochrone analysis
(Section~\ref{sub:star}), which yielded $2090_{-300}^{+270}$\,kg\,m$^{-3}$
and thus adds further credence to the hypothesis that both transiters are
orbiting the target star \TIC, rather than some unresolved companion.

\subsection{Transit timing variations}
\label{sub:TTVs}

\begin{figure*}
\begin{center}
\includegraphics[width=2.05\columnwidth,angle=0,clip=true]{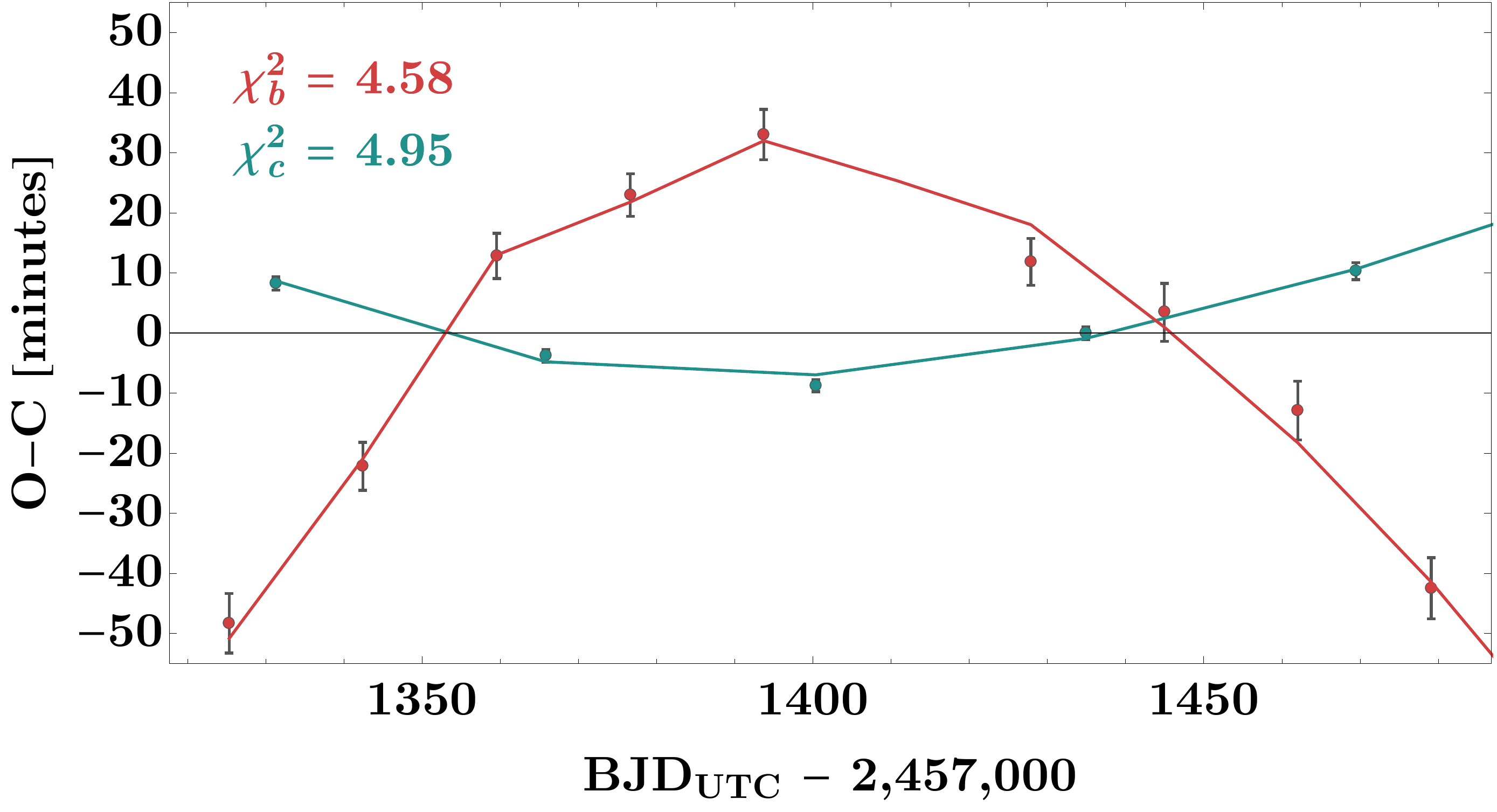}
\caption{\emph{
Observed minus calculated (O-C) transit times of TOI-216.01 (blue) and
TOI-216.02 (red). The TTVs are clearly anti-correlated indicating that
the transiters orbit the same primary. We overlay the maximum a-posteriori
dynamical model with solid lines, described in Section~\ref{sub:TTVs}.
}}
\label{fig:TTVs}
\end{center}
\end{figure*}

We plot the TTVs in Figure~\ref{fig:TTVs}, where one can clearly see the
strong case for anti-correlation mentioned earlier in this work. This
establishes that the transiters orbit the same star \citep{steffen:2013},
although this point alone does not establish the planetary nature of the
two transiters. Critically, their masses could potentially be consistent
with a brown dwarf or a late-type star, especially for TOI-216.01 whose
radius is similar to Jupiter.

The TTVs may be modeled by considering two masses orbiting a primary
with an N-body integrator. Dynamical analysis of the observed transit times
was performed with an symplectic N-body integrator code described in
\citet{nesvorny:2012}. The code was instructed to simultaneously fit all
transit times of both transiters, using \multi\ to perform the regression.
The integration time step was initially set to 0.6\,days, but we
also repeated for the final fits movig to a higher resolution of 0.3\,days
to ensure we recover consistent results - which indeed we do. We also
notew that the results are consistent to that derived when considering just
sectors 1-4, which was done in a previous draft of this paper.

Our dynamical model has 14 parameters: mass ratios $M_{\innerp}/M_{\star}$ and
$M_{\outerp}/M_{\star}$, orbital periods $P_{\innerp}$ and $P_{\outerp}$,
eccentricities $e_{\innerp}$ and $e_{\outerp}$, longitudes of periapsis
$\varpi_{\innerp}$ and $\varpi_{\outerp}$, impact parameters $b_{\innerp}$ and
$b_{\outerp}$, difference in nodal longitudes $\Omega_{\outerp} - 
\Omega_{\innerp}$, stellar density $\rho_{\star}$, and reference epochs
$\tau_{\innerp,\mathrm{ref}}$ and $\tau_{\outerp,\mathrm{ref}}$ between a
reference time and the first observed transit of each planet. All orbital elements
are given at the reference time 2,457,000\,$BJD_{\mathrm{UTC}}$ (TBJD).

We used uniform priors for all parameters except for $b_{\innerp}$,
$b_{\outerp}$ and $\rho_{\star}$, since our earlier light curve fits provide
strong constraints which can be leveraged here. Since \multi\ requires
simple parametric forms of the priors for the purposes of inverse transform
sampling, we approximated the marginal posteriors from our earlier fits
such that $b_{\outerp}$ is uniform between 0 and 0.4, $b_\innerp$ is a
Gaussian centered on 0.95 with 0.025 standard deviation and $\rho_{\star}$
is a Weibull prior with shape parameters $22.7$ and $2425.1$. These
priors are listed in Table~\ref{tab:priors2}.

\input{priors2_table.tex}

\begin{figure*}
\begin{center}
\includegraphics[width=2.05\columnwidth,angle=0,clip=true]{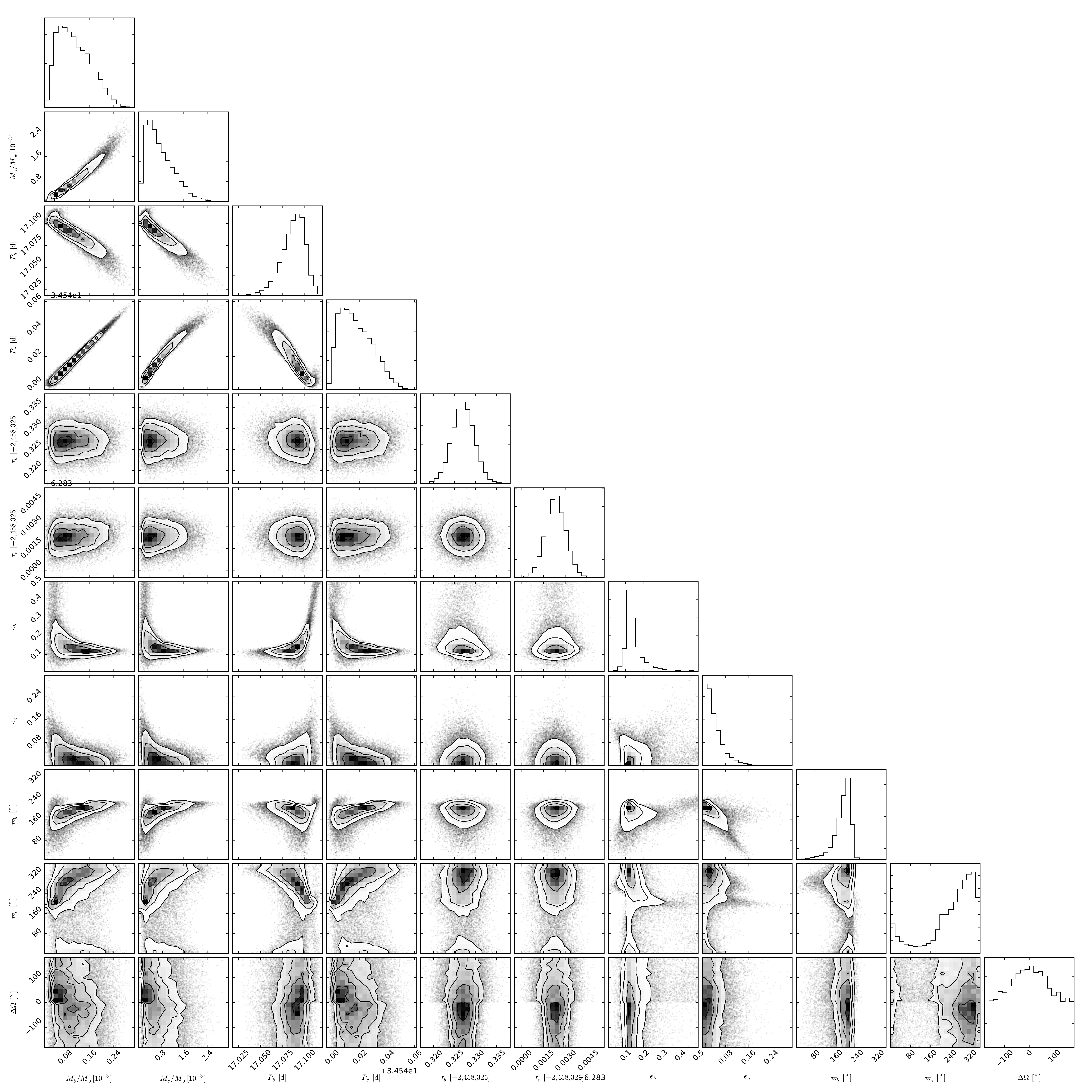}
\caption{\emph{
Corner plot of the joint posteriors from our dynamical fits to the
TTVs for the TOI-216 system. Since the inferred masses are planetary,
we dub the planets ``b'' and ``c'' here and in what follows. We omit
the terms with informative priors from the earlier light curve fits.
}}
\label{fig:corner}
\end{center}
\end{figure*}

The fits converged to a unique solution and the joint posteriors are depicted
in Figure~\ref{fig:corner} for reference. The maximum a-posteriori solution
is plotted in Figure~\ref{fig:TTVs}, which shows how the model is able to
fully describe the observed deviations.

Combining the derived mass ratios with the stellar mass derived earlier
(see Section~\ref{sub:star}) allows us to measure that 
$M_{\innerp} = 30_{-14}^{+20}$\,$M_{\oplus}$ and
$M_{\outerp} = 200_{-100}^{+170}$\,$M_{\oplus}$, which establishes that the
masses are far below the deuterium burning limit and these objects may
be classified as ``planets''. Accordingly, in what follows, we refer to
TOI-216.02 as TOI-216b (the inner planet), and TOI-216.01 as TOI-216c
(the outer planet).

\subsection{Final parameters}
\label{sub:finalparams}

To complete our analysis, we combine the fundamental stellar parameters
derived earlier (see Section~\ref{sub:star}) with the relative radii
(from Section~\ref{sub:fits}) and relative masses (from Section~\ref{sub:TTVs})
to calculate physical properties for both planets. Our final planet properties
are listed in Table~\ref{tab:final}.

\input{final_table.tex}

%% file: priors_table.tex
\begin{table}
\caption{
\emph{
Prior probability distributions adopted for the light curve fits. The syntax
$\mathcal{U}[a,b]$ denotes a continuous uniform distribution between real
values $a$ and $b$.
}
}
\centering 
\begin{tabular}{ll} 
\hline
parameter & adopted prior \\ [0.5ex] 
\hline
$R_1/R_{\star}$ & $\mathcal{U}[0,1]$ \\
$R_2/R_{\star}$ & $\mathcal{U}[0,1]$ \\
$b_1$ & $\mathcal{U}[0,2]$ \\
$b_2$ & $\mathcal{U}[0,2]$ \\
$q_1$ & $\mathcal{U}[0,1]$ \\
$q_2$ & $\mathcal{U}[0,1]$ \\
$\log_{10}$($\rho_{\star}$\,[kg\,m$^{-3}$]) & $\mathcal{U}[0,6]$ \\
$\tau_{\innerp,1}$\,[TBJD] & $\mathcal{U}[1324.335632,1326.335632]$ \\
$\tau_{\innerp,2}$\,[TBJD] & $\mathcal{U}[1341.434774,1343.434774]$ \\
$\tau_{\innerp,3}$\,[TBJD] & $\mathcal{U}[1358.533916,1360.533916]$ \\
$\tau_{\innerp,4}$\,[TBJD] & $\mathcal{U}[1375.633058,1377.633058]$ \\
$\tau_{\innerp,5}$\,[TBJD] & $\mathcal{U}[1392.732200,1394.732200]$ \\
$\tau_{\innerp,7}$\,[TBJD] & $\mathcal{U}[1426.930484,1428.930484]$ \\
$\tau_{\innerp,8}$\,[TBJD] & $\mathcal{U}[1444.029626,1446.029626]$ \\
$\tau_{\innerp,9}$\,[TBJD] & $\mathcal{U}[1461.128768,1463.128768]$ \\
$\tau_{\innerp,10}$\,[TBJD] & $\mathcal{U}[1478.227910,1480.227910]$ \\
$\tau_{\outerp,1}$\,[TBJD] & $\mathcal{U}[1330.285130,1332.285130]$ \\
$\tau_{\outerp,2}$\,[TBJD] & $\mathcal{U}[1364.824472,1366.824472]$ \\
$\tau_{\outerp,3}$\,[TBJD] & $\mathcal{U}[1399.363814,1401.363814]$ \\
$\tau_{\outerp,4}$\,[TBJD] & $\mathcal{U}[1433.903156,1435.903156]$ \\
$\tau_{\outerp,5}$\,[TBJD] & $\mathcal{U}[1468.442498,1470.442498]$ \\ [1ex]
\hline 
\label{tab:priors}
\end{tabular}
\end{table}

%% file: times_table.tex
\begin{table}
\caption{
\emph{
Medians and one-sigma uncertainties for the ten times of transit minimum in our light
curve fit of TOI-216.01 and TOI-216.02.
}
}
\centering 
\begin{tabular}{lll} 
\hline
parameter & epoch & $BJD_{\mathrm{UTC}}$ - 2,457,000 \\ [0.5ex] 
\hline
$\tau_{\innerp,1}$ & 1 & $1325.3277 \pm 0.0033$ \\
$\tau_{\innerp,2}$ & 2 & $1342.4306 \pm 0.0027$ \\
$\tau_{\innerp,3}$ & 3 & $1359.5398 \pm 0.0026$ \\
$\tau_{\innerp,4}$ & 4 & $1376.6316 \pm 0.0025$ \\
$\tau_{\innerp,5}$ & 5 & $1393.7234 \pm 0.0029$ \\
$\tau_{\innerp,7}$ & 7 & $1427.8784 \pm 0.0027$ \\
$\tau_{\innerp,8}$ & 8 & $1444.9574 \pm 0.0034$ \\
$\tau_{\innerp,9}$ & 9 & $1462.0308 \pm 0.0034$ \\
$\tau_{\innerp,10}$ & 10 & $1479.0951 \pm 0.0035$ \\ [0.5ex]
\hline
$\tau_{\outerp,1}$ & 1 & $1331.28509 \pm 0.00076$ \\
$\tau_{\outerp,2}$ & 2 & $1365.82443 \pm 0.00074$ \\
$\tau_{\outerp,3}$ & 3 & $1400.36868 \pm 0.00070$ \\
$\tau_{\outerp,4}$ & 4 & $1434.92243 \pm 0.00072$ \\
$\tau_{\outerp,5}$ & 5 & $1469.47729 \pm 0.00098$ \\ [1ex]
\hline 
\label{tab:times}
\end{tabular}
\end{table}

%% file: priors2_table.tex
\begin{table}
\caption{
\emph{
Prior probability distributions adopted for the N-body fits to the
transit times. The syntax $\mathcal{U}[a,b]$ denotes a continuous uniform
distribution between real values $a$ and $b$, $\mathcal{N}[a,b]$ is a normal
distribution with a mean of $a$ and variance $b^2$, and $\mathcal{W}[a,b]$ is
a Weibull distribution with scale parameter $a$ and shape parameter $b$.
The upper mass cut-off corresponds to approximately 4.6 Jupiter masses.
}
}
\centering 
\begin{tabular}{ll} 
\hline
parameter & adopted prior \\ [0.5ex] 
\hline
$M_{\innerp}/M_{\star}$ & $\mathcal{U}[0.0,0.005]$ \\
$M_{\outerp}/M_{\star}$ & $\mathcal{U}[0.0,0.005]$ \\
$P_{\innerp}$\,[days] & $\mathcal{U}[17.0,17.2]$ \\
$P_{\outerp}$\,[days] & $\mathcal{U}[34.3,34.7]$ \\
$\tau_{\innerp,\mathrm{ref}}$\,[TBJD] & $\mathcal{U}[1325.23 1325.43]$ \\
$\tau_{\outerp,\mathrm{ref}}$\,[TBJD] & $\mathcal{U}[1331.18 1331.38]$ \\
$e_{\innerp}$ & $\mathcal{U}[0,0.5]$ \\
$e_{\outerp}$ & $\mathcal{U}[0,0.5]$ \\
$\varpi_{\innerp}$\,[rads] & $\mathcal{U}[0,2\pi]$ \\
$\varpi_{\outerp}$\,[rads] & $\mathcal{U}[0,2\pi]$ \\
$b_{\innerp}$ & $\mathcal{N}[0.95,0.025]$ \\
$b_{\outerp}$ & $\mathcal{U}[0,0.4]$ \\
$\Omega_{\outerp} - \Omega_{\innerp}$\,[rads] & $\mathcal{U}[0,2\pi]$ \\
$\rho_{\star}$\,[kg\,m$^{-3}$] & $\mathcal{W}[2425,23]$ \\ [1ex]
\hline 
\label{tab:priors2}
\end{tabular}
\end{table}

%% file: final_table.tex
\begin{table}
\caption{
\emph{
Medians and one-sigma uncertainties for the system parameters of the planets
TOI-216b \& c. Note that $T_{14}$ is the first-to-fourth contact transit duration,
$T_{23}$ is the second-to-third contact transit duration,
$\tilde{T}$ is the transit duration from the planet's center entering to
exiting the stellar disk, $S$ denotes insolation and $T_{\mathrm{eq}}$ is the
equilibrium temperature assuming a zero-albedo blackbody.
}
}
\centering 
\begin{tabular}{lll} 
\hline
parameter & TOI-216b & TOI-216c \\ [0.5ex] 
\hline
$P$\,[days] & $17.089_{-0.015}^{+0.011}$ & $34.556_{-0.010}^{+0.014}$ \\ 
$\tau_0$\,[BJD$_{\mathrm{UTC}}-2,457,000$] & $1325.3270_{-0.0026}^{+0.0026}$ & $1331.28531_{-0.00067}^{+0.00068}$ \\ 
$b$ & $0.948_{-0.017}^{+0.027}$ & $0.15_{-0.10}^{+0.11}$ \\ 
$a/R_{\star}$ & $33.25_{-0.65}^{+0.46}$ & $53.18_{-1.04}^{+0.74}$ \\ 
$p$ & $0.0833_{-0.0082}^{+0.0168}$ & $0.1235_{-0.0014}^{+0.0014}$ \\ 
$\rho_{\star}$\,[kg\,m$^{-3}$] & $2380_{-140}^{+100}$ & $2380_{-140}^{+100}$ \\ 
$i$\,[$^{\circ}$] & $88.364_{-0.068}^{+0.042}$ & $89.83_{-0.12}^{+0.11}$ \\ 
$q_1$ & $0.44_{-0.18}^{+0.24}$ & $0.44_{-0.18}^{+0.24}$ \\ 
$q_2$ & $0.24_{-0.11}^{+0.18}$ & $0.24_{-0.11}^{+0.18}$ \\ 
$T_{14}$\,[hours] & $2.062_{-0.070}^{+0.068}$ & $5.514_{-0.047}^{+0.052}$ \\ 
$\tilde{T}$\,[hours] & $1.25_{-0.37}^{+0.17}$ & $4.890_{-0.050}^{+0.054}$ \\ 
$T_{23}$\,[hours] & 0 & $4.267_{-0.067}^{+0.061}$ \\ 
$R_P$\,[$R_{\oplus}$] & $7.69_{-0.83}^{+1.62}$ & $11.29_{-0.42}^{+0.58}$ \\ 
$e$ & $0.132_{-0.023}^{+0.059}$ & $0.029_{-0.020}^{+0.037}$ \\ 
$\varpi$\,[$^{\circ}$] & $193_{-35}^{+20}$ & $275_{-113}^{+55}$ \\ 
$\Omega$\,[$^{\circ}$] & $270$ & $270_{-110}^{+110}$ \\ 
$M_P$\,[$M_{\oplus}$] & $30_{-14}^{+20}$ & $200_{-100}^{+170}$ \\ 
$\rho_P$\,[kg\,m$^{-3}$] & $340_{-180}^{+310}$ & $760_{-380}^{+660}$ \\ 
$a$\,[AU] & $0.1293_{-0.0051}^{+0.0067}$ & $0.2069_{-0.0082}^{+0.0107}$ \\ 
$S$\,[$S_{\oplus}$] & $25.9_{-1.7}^{+2.2}$ & $10.1_{-0.66}^{+0.85}$ \\ 
$T_{\mathrm{eq}}$\,[K] & $628_{-11}^{+13}$ & $497_{-8}^{+10}$ \\ [1ex] 
\hline 
\label{tab:final}
\end{tabular}
\end{table}

%% file: discussion.tex
We have demonstrated that the \TESS\ planetary candidates TOI-216.01 \& .02
must orbit the same primary star given their anti-correlated TTVs. The light
curve derived stellar density found by fitting both signals yields a value
almost precisely equal to the target star's density from an isochrone analysis,
establishing that the objects indeed orbit the target rather than a
contaminant. Finally, we have regressed an N-body dynamical model to the
observed TTVs to demonstrate that the masses of each body are far below
the deuterium burning limit making these bona fide ``planets''.

The TOI-216 planetary system displays some close similarities to the
Kepler-9 system \citep{holman:2010}, but is 1.6 magnitudes brighter in $V$.
In both cases, one finds low-density gas giants in a 2:1 mean motion resonance
orbiting a Sun-like star at similar periods ($\sim$20\,d and $40$\,d). To a
lesser degree, the system also resembles KOI-872 \citep{nesvorny:2012}. In both
of these cases, the MMR pair of planets are accompanied by a short-period
super-Earth and thus it is natural to wonder if perhaps TOI-216 may also be
accompanied by a small and currently unresolved terrestrial planet. We ran a
box-least squares search \citep{BLS:2002} for such a signal but find no
significant peaks with the available \TESS\ data.

The TTVs of TOI-216 are characterized by a super-period as the longitude
of conjunctions circulates with a timescale of $\mathcal{O}[10^3]$\,days.
Although TOI-216 will be monitored throughout the first year of \TESS\
observations, the super-period looks likely to exceed this baseline and
thus continuous monitoring from the ground in 2020 would greatly benefit
the determination of precise orbital elements. With the limited phase
coverage available at the time of writing, the masses quoted in this work
will surely be refined considerably in the future.

The resonance between the gas giants is consistent with dissipative processes
in disk-planet interaction during their presumably inward migration from
beyond the snow line \citep{crida:2008,havel:2011,cimerman:2018}. In the
Grand Tack hypothesis of the Solar System \citep{hansen:2009,walsh:2011},
Jupiter is thought to have opened up a gap, migrating slower
(type II) than Saturn (type I), which likely failed to fully open a gap.
This enabled Saturn to catch up to Jupiter, trapping the pair
in resonance when Jupiter was at $\sim1.5$\,AU, which reversed subsequent
migration. In the case of both Kepler-9 and TOI-216, the gas giants have
maximum a-posteriori masses similar to Saturn, and thus may have
failed to have opened up full gaps, causing them to type I migrate far
deeper. As a larger sample of such systems is found in the future, it will
be interesting to test if these giants tend to have sub-Jupiter masses in a
statistically significant manner, as would be expected under this hypothesis.

Disk migration is likely a more favorable scenario for the survival of
satellite systems around these giants than planet-planet scattering
\citep{gong:2013}. Further, the fact that TOI-216 is relatively bright means
that follow-up with larger facilities would be well suited to make a search
for a satellite system. Similarly, the low-density (high scale height),
deep transits and reasonably bright target star would make TOI-216 a potential
target for atmospheric characterization of gas giants in a cooler regime to
their hot-Jupiter counterparts. Further, like Kepler-9, the two planets
provide an opportunity for differential transit spectroscopy alleviating
systematic effects.